

\input phyzzx
\def\papersize{\hsize=35pc \vsize=52pc \hoffset=0.5pc \voffset=0.8pc
   \advance\hoffset by\HOFFSET \advance\voffset by\VOFFSET
   \pagebottomfiller=0pc
   \skip\footins=\bigskipamount \normalspace }

\nopubblock
\rightline{YITP/U-92-32}
\rightline{DUCTP-92-53}
\title{ Covariance Properties  of  Reflection Equation Algebras}
\author{P. P. Kulish\footnote*{Address after October 1992 :
St.Petersburg Branch of Steklov Mathematical Institute,
Fontanka 27, St.Petersburg, 191011, Russia.}}
\vskip-10pt
\address{Yukawa Institute for Theoretical Physics,
Kyoto University, Kyoto 606, Japan}
\author{R. Sasaki }
\address{Uji Research Center,
Yukawa Institute for Theoretical Physics, \break
Kyoto University, Uji 611, Japan}
\andaddress{Department of Mathematical Sciences,
University of Durham, \break
Durham, DH1 3LE, United Kingdom}
\abstract{The reflection equations (RE) are a consistent extension of
the Yang-Baxter
 equations (YBE) with an addition of one element, the so-called reflection
matrix or $K$-matrix.
For example, they describe the conditions for factorizable scattering
 on a half line
just like the YBE  give the conditions for factorizable
scattering on an entire line.
The YBE  were  generalized to define quadratic algebras,
\lq Yang-Baxter algebras\rq\ (YBA), which were used intensively  for the
discussion of  quantum groups.
Similarly, the RE
define quadratic algebras, \lq the reflection equation algebras\rq\ (REA),
which enjoy  various remarkable properties both new and inherited from the
YBA.
Here we focus on the various properties  of the
REA, in particular, the quantum group comodule properties,
generation of a series of new solutions by composing known solutions,
the extended REA and  the central elements, etc.
}
\endpage


\def \AF {Alekseev A. and Faddeev L., {\it Commun. Math. Phys.} {\bf 141}
(1991) 413.}
\def \AFST {Alekseev A., Faddeev L. and Semenov-Tian-Shansky M.,
 Preprint LOMI, E-5-91 (1991); (in [\rKulE], 148.)}
\def \AFSTV {Alekseev A., Faddeev L., Semenov-Tian-Shansky M. and Volkov A.,
Preprint CERN-TH-5981/91 (1991).}
\def \AGS {Alvarez-Gaum\'e L., Gomez C. and Sierra G., in
\lq\lq Physics and Mathematics of Strings."
 Brink L., Friedan D. and Polyakov A. (Eds.), World Scientific,
 Singapore, (1990).}
\def\Bax{Baxter~R.~J., {\it Exactly Solved Models in Statistical Mechanics}
Academic Press, (1982).}
\def \BB {Babelon O., {\it Commun. Math. Phys.} {\bf 139} (1991) 619
and in  [\rKulE], 159.}
\def  \Beth{Bethe~H.~A., {\it Z.~Physik \bf 71} (1931) 205.}
\def\CDSWZ{Chryssomalakos C., Drabant B., Schlieker M., Weich W. and Zumino B.,
preprint UCB92/03, (1992).}
\def \CWSSW {Carow-Watamura U., Schlieker M., Scholl M. and  Watamura S.,
 {\it Z. Phys. C} {\bf 48} (1990) 159.}
\def \Che {Cherednik I.V., {\it Teor. Matem. Fiz.} {\bf 61} (1984) 55.}
\def \Chea {Cherednik I.V., preprint, Bonn-HE-90-04(1990);
in: Proc.RIMS-91-Project
\lq\lq Infinite analysis", Ed. T.Miwa (1992).}
\def \CFFS {Corrigan E., Fairlie D., Fletcher P. and Sasaki R.,
{\it J. Math. Phys.} {\bf 31} (1990) 776.}
\def \CG {Cremmer E. and Gervais J.-L.,
 {\it Commun. Math. Phys. \bf 144} (1992) 279; (in [\rKulE], 259.)}
\def \Dria {Drinfeld V.G., \lq\lq Quantum groups."
In: Proc.~ICM-86, {\bf 1}, Berkeley, AMS (1987) 798.}
\def \FRT {Faddeev L., Reshetikhin N. and Takhtajan L., {\it Alg. Anal.}
 {\bf 1} (1989) 178 (in Russian, English translation: Leningrad Math. J.
{\bf 1} (1990) 193).}
\def \FMai{Freidel L. and Maillet J.M., {\it Phys. Lett.} {\bf B262} (1991)
278; {\bf B263} (1991) 403.}
\def \Jim {Jimbo M., {\it Lett. Math. Phys.} {\bf 10} (1985) 63,
{\bf 11} (1986) 247; \nextline
{\it Commun. Math. Phys.} {\bf 102} (1986) 537.}
\def \Jima {Jimbo M. (Ed.),
\lq\lq Yang-Baxter equation in integrable systems."
World Scientific, Singapore (1990).}
\def \KSS {Kulish P., Sasaki R. and Schwiebert C.,  {\it J.~Math.~Phys.}
in press.}
\def \KulE {Kulish P.P. (Ed.), Proc.
\lq\lq  Euler Inter. Math. Inst. on Quantum Groups." {\it Lect. Notes Math.}
 {\bf 1510} Springer, Berlin (1992).}
\def \Kulg {Kulish P.P., ~Kyoto preprint YITP/K-959 (1991).}
\def \Kuli {Kulish P.P., ~Kyoto preprint YITP/K-984, (1992).}
\def \Kuz {Kuznetsov V.~B., {\it J.~Math.~Phys. \bf 31} (1990) 1167.}
\def \KoU {Kobayashi T. and Uematsu T., Kyoto preprint KUCP-47, (1992).}
\def \Majb {Majid S., {\it Int. J. Mod. Phys.} {\bf A5} (1990) 1;
{\it  J. Math. Phys.} {\bf 32} (1991) 3426; in [\rKulE];
preprint DAMTP/92-12.}
\def \Majd {Majid S., preprint DAMTP/92-65, (1992).}
\def \Manb {Manin Yu.I.,
\lq\lq Topics in noncommutative geometry." Princeton Univ. Press, (1991);
preprint MPI/91-60 (1991).}
\def \MNa {Mezincescu L. and Nepomechie R., {\it J. Phys. \bf A25} (1992)
2533;}
\def \MNb {in `Quantum Groups' Curtright T., Fairlie D. and
Zachos C. (Eds.), World Scientific (1991).}
\def \Nou{ Noumi M., preprint UT-Komaba (1991).}
\def \Ons {Onsager L., {\it Phys. Rev. \bf 65} (1944) 117.}
\def\OW{Ogievetsky E.,  Reshetikhin N. and  Wiegmann P., {
\it Nucl. Phys.} {\bf B280} (1987) 45.}
\def \Skla {Sklyanin~E.~K., {\it Sov. Phys. Dokl \bf 24} (1979) 107.}
\def \Sklc {Sklyanin E.K., {\it J. Phys. A. : Math. Gen.}
{\bf 21} (1988) 2375.}
\def \Skld {Kulish P.P. and Sklyanin E.K., {\it J. Phys. A.} to be published. }
\def \Sos {Sossinsky A.B., in [\rKulE].}
\def \SWZ {Schirrmacher A., Wess J. and Zumino B., {\it Z. Phys. C}
{\bf 49} (1991) 317.}
\def \SWZu{Schupp~P., Watts~P. and Zumino~B., preprint LBL-32314,
UCB-PTH-92/13, (1992).}
\def \Takha {Takhtajan L.A.,  in the
Proceedings of the Taniguchi Symposium at RIMS, (1988).}
\def \Tha {Thacker~H.~B., {\it Rev. Mod. Phys. \bf 53} (1981) 253.}
\def \WDA {Wadati M., Deguchi T. and Akutsu Y., {\it Phys. Reports}
{\bf C180} (1989) 427.}
\def \WZu {Wess J. and Zumino B., {\it Nucl. Phys.B} (Proc. Suppl.)
{\bf 18B} (1990) 302.}
\def \YGe {Yang C.N. and Ge M.-L. (Eds.),
\lq\lq Braid group, knot theory and statistical mechanics"
World Scientific, Singapore (1989).}
\def \YY {Yang~C.~N. and Yang~C.~P., {\it Phys. Rev. \bf 150}  (1966) 321, 327,
{\bf 151} (1966) 258.}
\def \Zuma {Zumino B.,
Preprint UCB-PTH-62/91 (1991).}
\def\Z{Zamolodchikov}
\def \ZZ {\Z ~A.~B.  and  \Z ~Al.~B, {\it Ann. Phys.} {\bf 120} (1979)
253.}



\def\w{\omega}

\def\pe{\ifmmode {\cal P} \else {${\cal P}$} \fi}
\def\CA{\ifmmode {\cal A} \else {${\cal A}$} \fi}
\def\CC{\ifmmode {\cal C} \else {${\cal C}$} \fi}
\def\CD{\ifmmode {\cal D} \else {${\cal D}$} \fi}
\def\CK{\ifmmode {\cal K} \else {${\cal K}$} \fi}
\def\CR{\ifmmode {\cal R} \else {${\cal R}$} \fi}
\def\Rt{\ifmmode {\widetilde R} \else {$\widetilde R$} \fi}
\def\At{\ifmmode {\widetilde A} \else {$\widetilde A$} \fi}
\def\Bt{\ifmmode {\widetilde B} \else {$\widetilde B$} \fi}
\def\Ct{\ifmmode {\widetilde C} \else {$\widetilde C$} \fi}
\def\Rti{\ifmmode {\widetilde{R\sp{-1}}} \else {$\widetilde{R\sp{-1}}$} \fi}
\def\wt{\widetilde}
\def\al{\alpha}
\def\be{\beta}
\def\ga{\gamma}
\def\de{\delta}
\def\ep{\varepsilon}
\def\NPrefs{\let\refmark=\NPrefmark}
\NPrefs

\def\to{\rightarrow}
\def\br{\hfil\break}

\chapter{Introduction}
Many remarkable  properties of the Yang-Baxter equations (YBE)
and Yang-Baxter Algebras (YBA), in particular, the fruitful application
of the $R$-matrix approach to the quantum group theory
\Ref\rFRT{\FRT},
 have prompted the study of
natural extensions of these concepts.
The reflection equations (RE)
appeared in the factorized scattering on a half line
\Ref\rChe{\Che}, and its algebraic structure, the reflection equation algebras
(REA) and various c-number solutions are studied in our previous papers
\REF\rSkld{\Skld}\REF\rKSS{\KSS}\refmark{\rSkld,\rKSS}.
The RE and REA appeared in papers dealing with quite different topics:
in quantum groups and quantum algebras,
in generalizing the concept of commuting transfer matrices to
 non-periodic boundary conditions on lattice models
\REF\rSklc{\Sklc}\REF\rMN{\MNa~~\MNb}\refmark{\rSklc,\rMN},
in modified Knizhnik-Zamolodchikov equations \Ref\rChea{\Chea},
in non-commutative differential geometry
\REF\rZuma{\Zuma}\REF\rNCDG{\AF\nextline\CWSSW\nextline\SWZ
\nextline\KoU\nextline\SWZu}
\refmark{\rZuma,\rNCDG},
in quantum Liouville theory
\Ref\rCG{\CG},
in the lattice Kac-Moody algebras
\REF\rAFST{\AFST\nextline\AFSTV}\REF\rBB{\BB}\REF\rFMai{\FMai}
\refmark{\rAFST,\rBB,\rFMai},
in quantum homogeneous spaces
\Ref\rNou{\Nou},
in integrable finite dimensional systems
\Ref\rKuz{\Kuz},
in the generalization of the braid group to manifolds of non-trivial
topology
\REF\rKulg{\Kulg}
\refmark{\rKulg,\rKSS} and in \lq\lq braided groups\rq\rq
\Ref\rMAJ{\Majb} etc.

In this paper we discuss various properties of the quadratic algebras
related with reflection equations (REA).
The main purpose is an extension of REA by generalizing the coincidence
of RE with the additional relation of the braid group of solid
handlebody (2.23). Namely the $K$-matrix of RE corresponds to $\tau$ (2.24),
an additional generator of the braid group of solid handlebody which
indicates the path of the first string going around the \lq\lq hole\rq\rq.
For a manifold of higher genus ($g>1$) we have braid group generators
$\tau$, $\tau'$ etc., corresponding to different \lq\lq holes\rq\rq.
They, in turn, correspond to $K$ , $K'$ etc. which generate two (many)
copies of REA's. The many copies of REA's thus generated do not commute
with each other because of the consistency conditions under the
quantum group coaction. The consistent relationships
among $K$ and $K'$ etc. together with the construction of new solutions
satisfying the REA by composing various known solutions constitute the
main content of the extended REA.
The central elements of the (extended) REA are derived in full generality
and their properties are discussed.
The basic idea is the covariance under the associated
quantum group coaction.
In other words we treat the REA as typical examples of
the \lq\lq quantum group tensors\rq\rq. This makes it easy to generalize the
concepts and applications of REA to a wider class of objects and algebras,
which will be discussed in some detail.

The present paper is organized as follows. In section 2 we give a
short description of the fields in which the YBE and YBA are useful.
It is pointed out that the  RE and related quadratic algebras have the same
range of applications. This section is for  setting a  proper stage for the
subsequent discussion and for introducing appropriate notation.
In section 3 we treat the \lq\lq extended REA\rq\rq, namely the multiple
copies of the REA as mentioned above, from the view point of the braid
group for higher genus manifolds.
In section 4 we discuss the quantum group tensors and give various examples.
We show that many properties of REA and its related algebras can be understood
 lucidly in this framework, in particular those discussed in section 3.
In section 5 the central elements of the REA and the
\lq\lq extended REA\rq\rq\
are discussed.
In section 6 we discuss the \lq\lq dual\rq\rq\ of the RE, which helps to
generalize  the concept of the commuting transfer matrix for non-periodic
 boundary conditions on  lattice statistical models.
The \lq\lq quantum group tensor \rq\rq\ viewpoint again gives a lucid
understanding of this concept and its generalization. The r\^ole played
by non-trivial c-number solutions of RE is emphasized here.
Section 7 contains a  summary and comments.

\chapter{Summary of Yang-Baxter equations (algebras) and Reflection equation
(algebras)}

The reflection equations (RE) and reflection equation algebras (REA) are
extension of the Yang-Baxter equation (YBE) and Yang-Baxter algebras (YBA),
which play very important r\^oles in solving various problems
in many branches of theoretical and
mathematical physics.

\section{YBE and YBA:\quad Summary}
Let us briefly review these well known subjects for the purpose of setting
a suitable stage for the discussion of RE and REA. We will mainly follow the
notation of our previous paper
\refmark\rKSS.

The YBE and YBA appeared in

 i) One dimensional quantum chains. \br
For example the Heisenberg spin
chain as investigated by Bethe
\Ref\rBeth{\Beth}\ and the \lq\lq Bethe ansatz\rq\rq\ for
determining the energy eigenvalues.

ii) Factorizable scattering in $1+1$ dimensions.\br
Here the YBE appears as the conditions for factorizability on an entire
line.
Many explicit examples are known \Ref\rYZOW{See for example, \YY
\nextline \Skla \nextline \Tha \nextline \ZZ \nextline \OW}.

iii) Statistical lattice models in 2 dimensions. \br
For example, the Ising model and various vertex models and others
\Ref\rBax{\Ons \nextline \Bax}.
The existence of the commuting transfer matrices is one of
the key points in this context.

iv) Braid groups.\br
The interaction among YBA, conformal field theory, braid group and knot theory
has produced quite fruitful results
\REF\rJima{\Jima} \REF\rYGe{\YGe}\REF\rAGS{\AGS}\REF\rWDA{\WDA}
\refmark{\rJima,\rYGe,\rAGS,\rWDA}.

Let us introduce various notions by using the language of  factorized
scattering.
By the assumption of the existence of an infinite set of quantum conserved
quantities all inelastic processes are forbidden. The set of outgoing
momenta is the same as the set of incoming momenta. The S-matrix element for
an $M$ particle elastic scattering is factorized into a product of
$M(M-1)/2$ two-particle  S-matrix elements.
In particular, for a three body S-matrix, there are two ways of
factorization (Fig.~1) which should be equal. This equality gives rise to
the Yang-Baxter equation.
The incoming or outgoing particles are  in one of the $N$-different internal
states specified by $i$ and $j$, $1\leq i,j\leq N$.
The energy-momenta of the relativistic
particles are conveniently parameterized by
the \lq rapidity\rq\ $p_a=m(\cosh\theta_a,\sinh\theta_a)$ thanks to the
two dimensionality. Here $\theta_a$ is the rapidity of the
$a$-th particle ($a=1,2,3$).
 The mass $m$ is assumed to be the  same for all the particles.
The internal state index $i$ and $j$ can change
after the scattering but the rapidities cannot due to the elasticity.
The S-matrix $R_{ab}$, $a,b=1,2,3$, being an $N\sp2\times N\sp2$ matrix of
the internal indices, depends
only on the difference of the rapidities $\theta_a -\theta_b$ because of
 the Lorentz invariance.
The Yang-Baxter equation corresponding to Fig.~1 can be expressed symbolically
as
$$ R(\theta_1-\theta_2)_{12}R(\theta_1-\theta_3)_{13}R(\theta_2-\theta_3)_{23}=
 R(\theta_2-\theta_3)_{23}R(\theta_1-\theta_3)_{13}R(\theta_1-\theta_2)_{12}.
\eqn\YBsim
$$
\def\stroke{\vrule height8pt width0.4pt depth-0.1pt}
\def\Ctext{{\rlap{\rlap{C}\kern 3.8pt\stroke}\phantom{C}}}
\def\C{\ifmmode{{\hbox\Ctext}}\else\Ctext\fi}
Namely YBE is a tensor equation acting on the tensor product of three
linear spaces $V_1\otimes V_2\otimes V_3$ ($V_1=V_2=V_3=\C\sp N$)
 and the suffices
indicate the spaces on which each S-matrix works.
If we write the matrix indices explicitly (suppressing the rapidities)
it reads
$$ {R_{i_1i_2~k_1k_2}}_{12}{R_{k_1i_3~j_1k_3}}_{13}{R_{k_2k_3~j_2j_3}}_{23}=
 {R_{i_2i_3~k_2k_3}}_{23}{R_{i_1k_3~k_1j_3}}_{13}{R_{k_1k_2~j_1j_2}}_{12},
\eqn\YBind
$$
in which the repeated indices ($k_1,k_2,k_3$) are summed, as usual.

Let us introduce another form of the YBE which is frequently used
in connection with the braid group
$$\hat R_{12}\hat R_{23}\hat R_{12} = \hat R_{23}\hat R_{12}\hat R_{23},
\eqn\hatYB
$$
in which $\hat R_{12} = \pe_{12}R_{12}$ and $\pe_{12}$ is the  permutation
operator of the first and the second spaces. Namely $\pe u\otimes v= v\otimes
u$. In terms of the indices it reads $\pe_{ij~kl} = \de_{il}\de_{jk}$.

There is only one step between the YBE and the YBA
\REF\rDria{\Dria} \REF\rJim{\Jim}
\refmark{\rDria,\rJim,\rFRT}.
We only have to consider the third space as a \lq quantum\rq\ space or
treat the indices in the third space as \lq invisible\rq
$$ R_{12}T_1T_2 = T_2T_1R_{12}, \eqn\FRTeq
$$
in which $T_1$ ($T_2$) is a matrix working on the first (second) space and a
linear operator working on the quantum space. This defines a quadratic algebra
generated by $T_{ij}$, each is a linear operator on the quantum space.
The YBA has an important algebra homomorphism, called comultiplication
$$\triangle T_{ij} = \sum_k T_{ik}\otimes T_{kj}, \eqn\coml
$$
in which $\otimes$ means that the two $T$'s belong to different quantum spaces
or that they are \lq independent\rq\ physical operators thus
commuting with each other.
Alternatively one can express this by introducing a different symbol for the
second (first) quantum space operator
$$\triangle T_{ij} = \sum_k T_{ik}T_{kj}', \qquad [T_{ij}, T'_{kl}] =0.
$$
Obviously the coproduct can be applied an arbitrary number of times.
Let us apply it for $L$ times, $L$ being the lattice size of a lattice problem,
then \FRTeq\ reads
$$\eqalign{& R(\theta_1-\theta_2)_{12}\left[(TT'\cdots T\sp{(L)})(\theta_1)
\right]_1
\left[(TT'\cdots T\sp{(L)})(\theta_2)\right]_2  \cr
&~~~~=
\left[(TT'\cdots T\sp{(L)})(\theta_2)\right]_2\left[(TT'\cdots T\sp{(L)})
(\theta_1)\right]_1
R(\theta_1-\theta_2)_{12},\cr} \eqn\FRTLeq
$$
in which the \lq rapidity\rq dependence (in this context the
\lq spectral parameter\rq\ $\lambda$ might be preferred to the
\lq rapidity\rq\ but we stick to the convention) is shown explicitly.
By multiplying $R_{12}\sp{-1}$ from the left and by taking the trace
in the first and the second spaces we get
$$ [\, t(\theta_1), t(\theta_2)] =0, \qquad
t(\theta)= {\rm Tr}\left((TT'\cdots T\sp{(L)})(\theta)\right).
\eqn\comTra
$$
This gives the (one parameter ($\theta$) family of) commuting transfer matrix,
 which is the corner stone for the solvable lattice models.
It should be remarked that the trace obviously corresponds to the periodic
boundary condition in one direction of the lattice.
The generalization of the commuting transfer matrix for non-periodic
boundary conditions can be achieved in terms of REA and will be discussed
in section 6.

For most of the discussion in this paper we deal with the
\lq rapidity independent\rq\ Yang-Baxter equations and Yang-Baxter algebras.
Here is the simplest example of the solution of the
\lq rapidity independent\rq\ YBE associated with $GL_q(2)$ in the fundamental
representation,
$$
R=\pmatrix{q& & & \cr  &1& & \cr  &\w&1& \cr  & & &q\cr} , \quad
\quad \w=q-q^{-1} , \eqn\Rmat
$$
in which $q$ is a complex deformation parameter.
We adopt the convention
that any matrix elements not written explicitly are zeros.
Then the YBA \FRTeq\ determines a quadratic algebra generated by a $2\times2$
matrix $T$, which is the well known $GL_q(2)$ quantum group:
$$
T=\pmatrix{a&b \cr  c&d \cr} : \qquad
\eqalign{a b&=q b a , \cr  a c &=q  c a , \cr}  \qquad
\eqalign{ b d &=q d  b , \cr   c  d &=q  d  c  , \cr}  \qquad
\eqalign{a d -q b c &= d  a-q\sp{-1} b c  , \cr   b c &= c  b . \cr}
\eqn\Qgrop
$$

At the end of the review of  the YBE and YBA
let us discuss the braid group. The braid group with $n$ strands is
generated by $n-1$ generators $\{\sigma_i\}$
with the following set of relations (Fig.~2)
$$\eqalign{
\sigma_i\sigma_j &= \sigma_j\sigma_i , \qquad \vert i-j\vert\geq2 \cr
\sigma_i\sigma_{i+1}\sigma_i&=\sigma_{i+1}\sigma_i\sigma_{i+1}. \cr}
\eqn\BRrel
$$
These are the relations implied by the YBE \hatYB\ by the identification
$\sigma_i = \hat R_{i\,i+1}$.

\section{RE and REA:\quad Summary}
Here we briefly summarize the basic properties of the Reflection Equations
(RE) and the Reflection Equation Algebras (REA) in connection with YBE and YBA.

The most natural stage to extend the YBE to the RE  is the
factorizable scattering with reflection at the end of a line,
namely the factorizable scattering on a half line.
Then the factorizability conditions consist of two parts,
the first part is the YBE itself \YBsim\ and the second part contains an
additional element, called $K$-matrix, describing the reflection
at the endpoint of the half line
(Fig.~3)\refmark\rChe.
$$
R\sp{(1)}_{12}K_1R\sp{(2)}_{12}K_2 = K_2R\sp{(3)}_{12}K_1R\sp{(4)}_{12}.
 \eqn\REgen
$$
Here $K$ is an $N\times N$ matrix and the various (rapidity dependent)
$R$-matrices
$R\sp{(1)}$,$\ldots$,$R\sp{(4)}$  are related to the $R$-matrix in the
YBE and will be specified later.
For notational simplicity the rapidity dependence is suppressed.
(The suffices of the $R$-matrices like $R = R_{12}$
indicating the base space $V_1\otimes V_2$ will be suppressed in most cases.)
As in the YBE case we mainly discuss the RE and REA in the rapidity
independent form.

The above equation can be considered as a quadratic equation for
$N\times N$ c-number entries of $K$ \refmark\rKSS\ for a given solution
$R$ of YBE.
On the other hand they can be considered as the defining relations
of a quadratic algebra (Reflection Equation Algebra) generated by $K_{ij}$
just like the YBA.
Then the c-number solutions can be considered as the one-dimensional
representations of the algebra.

We require the REA to be closely related with the YBA, namely $K$ to be
a \lq\lq quantum group comodule\rq\rq. In other words the REA should be
invariant under either of the following transformations:
$$
K\to\delta(K)=\cases{&$TKT\sp t$ \cr  &$TKT\sp{-1}$\cr}
\qquad [K_{ij}, T_{kl}]=0.
\eqn\twococat
$$
This requirement leads to two groups of RE, called RE1 and RE2
\refmark{\rSkld,\rKSS} for which we write the standard
forms.
For RE1, namely for
$ \delta(\bar K)=T\bar KT\sp t$:
$$
R \bar K_1 R\sp{t_1} \bar K_2 = \bar K_2 {\widetilde {R\sp{t_1}}} \bar K_1
{\widetilde {R\sp{t_1\circ t_2}}},
\eqn\rethree
$$
and for RE2, namely for $ \delta(K)=TKT\sp{-1}$ :
$$
RK_1\Rt K_2 = K_2RK_1\Rt.  \eqn\REtw
$$
In order to distinguish these two groups of RE's we put a bar on the
$K$-matrix appearing in the RE1.
Here $t_1$ means transposition in the first space and $\wt A =\pe A\pe$ for
any matrix on $V_1\otimes V_2$. The other members of the group are obtained by
simply replacing any $R$ by $\Rt\sp{-1}$ \refmark\rSkld.
But we mainly discuss these forms of RE's.

Let us give the explicit forms of the  REA's for the $GL_q(2)$ $R$-matrix
given above \Rmat. For the RE1 :
$$
\bar K=\pmatrix{\al&\be\cr  \ga&\de\cr} \eqn\kreon
$$
 which satisfy the relations \refmark{\rSkld,\rKSS}
$$
\eqalign{[\al,\be]&=\w\al\ga , \cr  \al\ga&=q^2 \ga\al , \cr}  \qquad
\eqalign{[\al,\de]&=\w(q\be\ga+\ga^2) , \cr  [\be,\ga]&=0 , \cr}  \qquad
\eqalign{[\be,\de]&=\w\ga\de , \cr  \ga\de&=q^2 \de\ga .
\cr} \eqn\Ala
$$
 This algebra  has two central elements, linear and quadratic in $\bar K$,
$$ c_1 = \be - q\ga , \qquad c_2 = \al\de -q\sp2\be\ga . \eqn\qoncen
$$
The c-number solutions are
$$
\bar K=\ep_q=\pmatrix{ &1\cr  -q& \cr} , \quad
\bar K=\pmatrix{\al&\be\cr  &\de\cr} . \eqn\sol
$$
For the RE2 :
$$
K=\pmatrix{u&x\cr  y&z\cr} ,  \eqn\kretw
$$
then we find that the component form of the algebra  reads
\refmark{\rMAJ,\rKSS}
$$
\eqalign{ux&=q^{-2}xu , \cr  uy&=q^2yu , \cr}  \qquad
\eqalign{[u,z]&=0 , \cr  [x,y]&=q^{-1}\w (uz-u^2) , \cr}  \qquad
\eqalign{[x,z]&=-q^{-1}\w ux , \cr  [y,z]&=q^{-1}\w yu  .
\cr} \eqn\Alb  $$
The central elements of the above algebra  are also known
\refmark{\rKSS,\rZuma}
and they are invariant under the $GL_q(2)$ coaction
$$ c_1 = u + q\sp2 z , \quad c_2 =
uz -q\sp2 yx ,
 \eqn\centwo
$$
 The corresponding c-number solutions are
$$ K=\pmatrix{1& \cr &1\cr} , \quad K=\pmatrix{ & x \cr y & z
\cr} . \eqn\solt $$

It is interesting to note that the RE2, a generalization of the YBE,
has also an interpretation of the braid group for solid handlebody
(\ie,  genus 1)
\REF\rSos{\Sos}\refmark{\rKulg,\rMAJ,\rSos,\rKSS}.
Here we need to introduce one additional generator $\tau$, which
corresponds to a single turn around a hole (Fig.~4). As is shown in
Fig.~5 it satisfies a relation $$
\sigma_1\tau\sigma_1\tau = \tau\sigma_1\tau\sigma_1 .
\eqn\Breq
$$ This is just relation \REtw\ if we identify $$K_1 = \tau, \qquad
\hat R =\sigma_1.
\eqn\BRRid
$$


\chapter{Extended Reflection Equation Algebras}

The above relationship between the REA and the braid group for solid
handlebody ($g=1$) \BRRid\ can be readily generalized to a braid
group $B\sp{(g)}_n$ for a manifold of an arbitrary genus $g$.
The corresponding REA, to be called \lq\lq extended REA\rq\rq\
consists of $g$-copies of REA,
$ K\sp{(1)}, \ldots, K\sp{(g)}$,
$$  R K_1\sp{(j)} \Rt K_2\sp{(j)} = K_2\sp{(j)} R K_1\sp{(j)} \Rt ,
\qquad j=1,\ldots, g. \eqn\retwo
$$
The mutual relationship among various copies should also be covariant under
the \lq\lq quantum group comodule\rq\rq $K\to\delta(K)=TKT\sp{-1}$.
This induces non-trivial relations among them, since the trivial ones (\ie
simply commuting) are not preserved by the comodule
$$[\de(K\sp{(i)}_{st}), \de(K\sp{(j)}_{uv})]\neq 0,\quad {\rm even~ if}
\quad [K\sp{(i)}_{st}, K\sp{(j)}_{uv}]= 0.
$$
We are led to the following relations between $K\sp{(i)}$ and $K\sp{(j)}$
\REF\rCDSWZ{\CDSWZ}\refmark{\rMAJ,\rCDSWZ},
which  are \lq\lq linearly ordered'' (\ie\ non-symmetric in $i$ and $j$),
$$ K\sp{(i)} \succ K\sp{(j)},\quad i > j ,\qquad \Leftrightarrow \qquad
R\sp{-1}K_2\sp{(j)}R\,K_1\sp{(i)} = K_1\sp{(i)}R\sp{-1}K_2\sp{(j)}R.
 \eqn\Comp
$$
By keeping the \lq\lq order\rq\rq\ in the matrix multiplication we can
\lq\lq compose\rq\rq\
$K\sp{(i,j)}\equiv K\sp{(i)}K\sp{(j)}$,
 $i>j$
which also satisfies the REA,
$$  R K_1\sp{(i,j)} \Rt K_2\sp{(i,j)} = K_2\sp{(i,j)} R K_1\sp{(i,j)} \Rt .
\eqn\reij
$$
It should be remarked that the unit element $K=1$ always satisfies both
\retwo\ and \Comp.
The above \lq\lq composition rule'' and the \lq\lq linear order'' and most of
the composition rules below are easily understood pictorially
by the following identification $K\sp{(j)}\Leftrightarrow
\tau_j$ (Fig.~6,7,8) of the extended REA  with $B\sp g_n$ ($g\ge1$).
The above relation \Comp\ can be rewritten in an equivalent form
$$R K_1\sp{(i)}R\sp{-1}K_2\sp{(j)} = K_2\sp{(j)}R\,K_1\sp{(i)}R\sp{-1}.
\eqn\Compp
$$

The extended REA has several interesting algebraic properties.
 After the composition the  linear order is preserved.
Namely,
$$ K\sp{(i,j)}\succ K\sp{(k)}, \quad{\rm for}~ i>j>k .      \eqn\compre
$$
 This implies the second (and further) step of composition is possible.
Namely
$$K\sp{((i,j),k)}\equiv K\sp{(i,j)}K\sp{(k)} = (K\sp{(i)}K\sp{(j)})K\sp{(k)}
\eqn\compthr
$$
also satisfies the REA.
 The  linear order is preserved in the other direction, too
$$ K\sp{(i)}\succ K\sp{(j,k)}, \quad{\rm for}~ i>j>k ,      \eqn\comprd
$$
which results in  another composition
$$K\sp{(i,(j,k))}\equiv K\sp{(i)}K\sp{(j,k)} = K\sp{(i)}(K\sp{(j)}K\sp{(k)}).
\eqn\compthd
$$
 By the associativity we have simply
$$K\sp{((i,j),k)}= K\sp{(i,(j,k))} = K\sp{(i,j,k)} =
K\sp{(i)}K\sp{(j)}K\sp{(k)}, \quad i>j>k ,
 \eqn\compthdd
$$
satisfying the REA.

 A different type of composition is possible.
Let us define a notation
$$ K\sp{(-i)} \equiv (K\sp{(i)})\sp{-1}.              \eqn\Kinv
$$
Then
$$K\sp{(i,j,-i)}\equiv K\sp{(i)}K\sp{(j)}K\sp{(-i)},\quad
{\rm for}\quad i>j \eqn\compiji
$$
 also satisfies the REA.
 The linear order is preserved for this composition, too.
Namely
$$K\sp{(i,j,-i)}\succ K\sp{(k)}, \quad {\rm for}\quad i>j>k, \eqn\presKinv
$$
and
$$ K\sp{(i)}\succ K\sp{(j,k,-j)},\quad {\rm for}\quad i>j>k. \eqn\presKind
$$
That is $K\sp{(i)}K\sp{(j)}K\sp{(-i)}K\sp{(k)}$ and
$K\sp{(i)}K\sp{(j)}K\sp{(k)}K\sp{(-j)}$ satisfy the REA.

Yet another type of composition
is possible. Namely
$$K\sp{(-j,i,j)}\equiv K\sp{(-j)}K\sp{(i)}K\sp{(j)},\quad {\rm for}
\quad i>j \eqn\compjij
$$
 also satisfies the REA.
Correspondingly, the linear order is  preserved
for this composition, too
$$K\sp{(-j,i,j)}\succ K\sp{(k)}, \quad {\rm for}\quad i>j>k, \eqn\presKjnv
$$
and
$$ K\sp{(i)}\succ K\sp{(-k,j,k)},\quad {\rm for}\quad i>j>k. \eqn\presKjnd
$$
That is $K\sp{(-j)}K\sp{(i)}K\sp{(j)}K\sp{(k)}$ and
$K\sp{i)}K\sp{(-k)}K\sp{(j)}K\sp{(k)}$, $i>j>k$ satisfy the REA.

Here we have demonstrated various explicit examples of the composition
rules for the REA in the framework of the extended REA.
One could   discuss them from a slightly different point of view.
Namely in terms of  \lq\lq corepresentation'' and \lq\lq twisted (braided)
coproduct\rq\rq\  \refmark\rMAJ\ for REA.
Let $V$ be a linear space. We introduce
a \lq\lq corepresentation\rq\rq\ of REA $\wt\delta$:
$V\to K\otimes V$  in such a way
that the \lq\lq twisted (braided) coproduct
$\wt \triangle$\rq\rq\ for $K$ is consistent with  $\wt\delta$ ;
$$(\wt\triangle\otimes{\rm id})\circ\wt\delta =
({\rm id}\otimes\wt\delta)\circ\wt\delta,
\eqn\corep
$$
in which
$$ \wt\triangle(K)=K\,K\sp{'},\qquad \wt\delta(X) = KX,\quad (X\in V).
 \eqn\corepp
$$
Namely both sides of \corepp\ give $ (K\,K\sp{'})X=K\,(K\sp{'}X)=K\,K\sp{'}X$.
However, there seem to be many more  cases of compositions
than is naively suggested by the \lq\lq twisted (braided) coproducts\rq\rq\
\corepp.

Before closing this section, let us give a brief remark that the
\lq\lq extension\rq\rq\ and \lq\lq composition\rq\rq\  of the REA
can be generalized to a class of
associative algebras proposed in \refmark\rFMai.  Let us take, for simplicity,
the case corresponding to $g=2$. Namely we have two copies $K$ and $K'$
satisfying
$$\eqalign{  A K_1 \Ct K_2 &= K_2 C K_1 \At,  \cr
  A K_1\sp{'} \Ct K_2\sp{'} &= K_2\sp{'} C K_1\sp{'} \At. \cr}
\eqn\ABCDK
$$
 Here $A$ and $C$ should satisfy the YBE and its related equations
\refmark\rFMai.
(The special case $A=C=R$ reduces to the REA and to the extension arguments
given above.)
The relation between  $K$ and $K\sp{'}$  reads
$$ K\succ K'\qquad \Ct\sp{-1}K_1\sp{'}\Ct K_2 = K_2\At\sp{-1}K_1\sp{'}\Ct,
\eqn\comAB
$$
which can be rewritten as
$$ K_1A\sp{-1}K_2\sp{'}C  = C\sp{-1}K_2\sp{'}C K_1. \eqn\comABp
$$
Then as before  $KK\sp{'}$ satisfies the algebra relations
$$  A (KK\sp{'})_1 \Ct (KK\sp{'})_2 = (KK\sp{'})_2 C (KK\sp{'})_1 \At ,
\eqn\ABCDKKp
$$
It is interesting to check wether the other compositions of the extended
REA  work.

\chapter{Quantum Group Tensors }
At first sight the relations among $K\sp{(i)}$ and $K\sp{(j)}$ and their
composition rules discussed in the previous section look quite strange
algebraically and/or from the point of view of a physical model.
Here we show that such relations are rather common among the \lq\lq quantum
group tensors\rq\rq\ or \lq\lq $q$-tensors\rq\rq\ for short.
This is quite a general and useful concept and it contains
as a particular subclass
 the various relations of $q$-differential geometry (non-commutative
differential geometry).

In general terms tensors $A_{ij\ldots}\sp{st\ldots}$ are covariant and
 contravariant objects which transform
$$A_{ij\ldots}\sp{st\ldots}\to (\ell)_{ii'}(\ell)_{jj'}\cdots
(\ell\sp{-1})\sp{s's}(\ell\sp{-1})\sp{t't}\cdots A_{i'j'\ldots}\sp{s't'\ldots}.
$$
In this sense the $K$-matrices of REA are $q$-tensors
due to the comodule properties. Namely the $\bar K$ of RE1 is a rank (2,0)
$q$-tensor \refmark\rSkld
$$\bar K_{ij}\to \de(\bar K_{ij})=(T\bar KT\sp t)_{ij}= T_{ii'}T_{jj'}
\bar K_{i'j'}, \eqn\Twoten
$$
and $K$ for RE2 is a rank (1,1) $q$-tensor
$$ K_{ij}\to \de( K_{ij})=(T KT\sp{-1})_{ij}= T_{ii'}T\sp{-1}_{j'j} K_{i'j'}.
\eqn\Oneoneten
$$
An obvious difference from the ordinary tensors is that the transformation
coefficients $T_{ij}$ are not c-numbers but satisfy the YBA \FRTeq.

Let us  begin with some explicit examples of the consistent set of
$q$-tensors.
The first example is the  well known \lq\lq$q$-hyperplanes\rq\rq\ $\{x_i\}$,
$x_ix_j= q x_jx_i$ for $1\leq i<j\leq N$,
which can be written neatly in terms of the $GL_q(N)$ $R$-matrix in the
fundamental representations
$$ X= \{x_i\},\quad R X_1X_2 = qX_2X_1,
\quad {\rm or}\quad \hat R X_1X_2 = qX_1X_2, \eqn\oneten
$$
where $\hat R=\pe R$.
Another example are the \lq\lq one-forms\rq\rq\ on the
$q$-hyperplanes $\{\xi_i\}$,
$\xi_i\xi_j= -1/q \xi_j\xi_i$ for $1\leq i<j\leq N$.
They can also be written by using the same $R$-matrix as above
$$\Xi= \{\xi_i\},\quad R \Xi_1\Xi_2 =-1/ q\Xi_2\Xi_1,
\quad {\rm or} \quad \hat R \Xi_1\Xi_2 =-1/ q\Xi_1\Xi_2. \eqn\mioneten
$$
In these examples $q$ and $-1/q$ in the right hand side
are the eigenvalues of $\hat R$.
These relations are invariant under the $GL_q(N)$ group coaction
\Ref\rManb{\Manb}:
$$ \delta :  x_i \to T_{ij}x_j, \quad \xi_i \to  T_{ij}\xi_j,
  \eqn\xcom
$$
for which they are assumed to
commute with the elements of the quantum group matrix $T_{ij}$,
$[\,T_{ij}, x_k]= [\,T_{ij},\xi_k]=0$.
They are rank one $q$-tensors or \lq\lq $q$-vectors\rq\rq, for short.
 Now it is easy to give a general definition of the $q$-vector $V\sp{(1)}$
$$ V\sp{(1)} = \{v_i\},\quad
 R V_1\sp{(1)}V_2\sp{(1)} = \alpha V_2\sp{(1)}V_1\sp{(1)},
\quad {\rm or}\quad \hat R V_1\sp{(1)}V_2\sp{(1)} =
 \alpha V_1\sp{(1)}V_2\sp{(1)},
\eqn\defoneten
$$
in which $R$ is an arbitrary $R$-matrix and $\alpha$ is one of the
eigenvalues of $\hat R$.
In other words $V\sp{(1)}_1V\sp{(1)}_2$ lies in the eigenspace of $\hat R$
belonging to the eigenvalue $\al$.
As is clear from the above examples, we get different algebras (having the
same $q$-tensor properties) for different choices of eigenvalues of $\hat R$.
The dimension of the eigenspace with respect to the total space determines
the number of quadratic relations imposed by the condition \defoneten.
 The relations \defoneten\ are invariant under the quantum
group coaction specified by the $R$-matrix,
 $$ \delta :  v_i \to T_{ij}v_j.
  \eqn\vcom
$$
The relationships among various $q$-tensors are severely restricted
by the requirement of consistency with the quantum group coaction.
 The situation is similar to the case of the extension of the reflection
equation algebras
in which many copies of $K$ appear \retwo\ with nontrivial exchange
relations among them \Comp.
One of the simplest such examples are those among the two
$q$-vectors $X$ and $\Xi$ above
$$R\Xi_1X_2 = 1/qX_2\Xi_1, \quad {\rm or}\quad
 \hat R\Xi_1X_2 = (1/q)X_1\Xi_2.
\eqn\XXirel
$$
These are the relations
among the coordinates of the $q$-hyperplane and the corresponding one-forms,
see, for example,\Ref\rWZu{\WZu}.

One consistent exchange relation among two $q$-vectors,
say $V\sp{(1)}$ and ${V'}\sp{(1)}$, is obtained by generalizing
the above example:
$$\hat R{V'}\sp{(1)}_1V\sp{(1)}_2 =\be V\sp{(1)}_1{V'}\sp{(1)}_2,
\eqn\VVprrel
$$
in which $\be$ is an arbitrary constant (to be determined by some
requirements other than the quantum group covariance).
Take, for example, $N$ different copies of $q$-hyperplanes $X\sp{(j)}$
$(1\leq j\leq N)$, $X\sp{(j)}=\{ x\sp{(j)}_i, 1\leq i\leq N\}$
and assume that they satisfy the above consistency relations with $\be=1$
\REF\rKuli{\Kuli}\refmark{\rSkld,\rKuli}
$$\hat R X\sp{(j)}_1X\sp{(i)}_2 = X\sp{(i)}_1X\sp{(j)}_2, \quad 1\leq i< j
\leq N. \eqn\Nhyperp
$$
Then $U_{ij}=x\sp{(j)}_i$ satisfy exactly the same exchange relations as
$T_{ij}\in GL_q(N)$ and the coaction $\delta U_{ij}$ is equivalent to the
quantum group coproduct $\triangle T_{ij}$.
This is another characterization of the
quantum group itself in terms of many copies of $q$-tensors.

Another example of the exchange relations for two (many) copies of
$q$-vectors occurs in the following problem (see also \Ref\rMAJa{\Majd}):
to construct a $q$-plane as a sum of two
$q$-planes, $X\sp{(1)}={x\choose y}$, $X\sp{(2)}={w\choose z}$,
namely
$$ (x + w)(y + z) = q (y + z)(x + w) .
$$
This is achieved by taking $\be=1/q$ (which is one of the eigenvalues  of
the corresponding $\hat R$ with a minus sign)  in \VVprrel,
$$\hat R X\sp{(2)}_1X\sp{(1)}_2 =1/q X\sp{(1)}_1X\sp{(2)}_2,
\eqn\XXrel
$$
or, more explicitly,
$$ \eqalign{ xw = q\sp2 wx,\quad & xw=q(zx+\omega wy), \cr
             yw =qwy,\quad & yz=q\sp2zy.\cr}
$$
The same problem for the  one-forms is also conceivable,
namely to form a  one-form by
a sum of  two copies, $\Xi\sp{(1)}={\xi\choose\eta}$ and
$\Xi\sp{(2)}={\zeta\choose\mu}$,
$$ (\xi+\zeta)\sp2=(\eta+\mu)\sp2=0,\quad
q(\xi+\zeta)(\eta+\mu)+(\eta+\mu)(\xi+\zeta)=0.
$$
This can be achieved by taking $\be=-q$ (again one of the eigenvalues of
$\hat R$ with a minus sign) in \VVprrel,
$$\hat R \Xi\sp{(2)}_1\Xi\sp{(1)}_2 =-q \Xi\sp{(1)}_1\Xi\sp{(2)}_2.
$$
It is straightforward to generalize the above examples to the $GL_q(N)$
case.

It is straightforward to get a rank two $q$-tensor by taking a tensor
product of
two arbitrary rank one $q$-tensors.
If we define $\bar K_{ij} = v_iv'_j$ then
$\bar K$ is a rank two $q$-tensor
whose coaction is given by
$$ \bar V\sp{(2)} = \{\bar K_{ij}\}~{\rm for ~RE1},
\qquad  \delta : \bar K_{ij} \to
T_{il}T_{jm} \bar K_{lm}=(T\bar KT\sp t)_{ij}. \eqn\tworaten
$$
By combining \defoneten\ (both copies are assumed to have the same $\al$) and
\VVprrel\ (for arbitrary $\be$)
it is easy to show that $\bar K$ satisfies the RE1
$$R \bar K_1 R^{t_1}\bar K_2 =
\bar K_2\wt{R^{t_1}} \bar K_1 \wt{R\sp{t_1\circ t_2}}.
\eqn\reone
$$
In the $GL_q(N)$ example above,  we can also choose $\bar K=X\otimes \Xi$ or
$\bar K_{ij}=x_i\xi_j$, then it satisfies the RE1 with {\it an extra minus
sign} on the right hand side.

For rank two $q$-tensors (that is for (2,0) $q$-tensors)
one can introduce the notion of
\lq $q$-symmetry\rq\  and
\lq $q$-antisymmetry\rq\
\Ref\rCFFS{\CFFS} analogous to the ordinary tensor
calculus.
For concreteness, let us discuss $GL_q(N)$ (the fundamental representation)
 case.
The $q$-symmetric tensor $\bar K_S$ and a $q$-antisymmetric tensor are
$$ (\bar K_S)_{ij} = q(\bar K_S)_{ji}, \quad 1\leq i< j\leq N,
\eqn\defKS
$$
\vskip-20pt
$$ q(\bar K_A)_{ij} + (\bar K_A)_{ji}=0, \quad 1\leq i< j\leq N,
\quad (\bar K_A)_{ii} =0.\eqn\defKA
$$
It is straightforward to show that the above definitions of $\bar K_S$ and
$\bar K_A$ are consistent with the $GL_q(N)$ coaction.
The simplest examples are given by the $q$-hyperplanes and the one-forms,
$$(\bar K_S)_{ij}= x_ix_j,\quad (\bar K_A)_{ij}=\xi_i\xi_j.
$$

An arbitrary rank two tensor $\bar K$ can be uniquely decomposed into
a sum of a $q$-symmetric and $q$-antisymmetric parts.
This simply corresponds to the fact that the tensor product of
two fundamental representations can be decomposed into a direct
sum of the symmetric and the antisymmetric representations.
For example, $\bar K$ for $GL_q(2)$ (see \Ala)
is decomposed as
$$ \bar K= \pmatrix{\al &\be \cr \ga & \de}=\bar K_S + \bar K_A,
$$
\vskip-10pt
$$\bar K_S =\pmatrix{\al &qS \cr S &\de\cr},\quad
\bar K_A =\pmatrix{ &1\cr -q &\cr}A =\epsilon_qA,
$$
\vskip-10pt
$$S={1\over 1+q\sp2}(q\be+\ga),\quad A={1\over 1+q\sp2}(\be-q\ga)=
{1\over 1+q\sp2}c_1,
$$
in which $c_1$ is the central element of  the REA \qoncen.
Thus in this simplest example,
the antisymmetric part commutes with the symmetric part.
Under the requirement of the vanishing antisymmetric part ($A=c_1=0$)
the symmetric part
is related with the quantum homogeneous space.

As remarked above
the $K$-matrices of REA are  rank two (to be more precise a (2,0) and
a (1,1) rank) $q$-tensors
since their comodule properties are given by\Twoten\ and \Oneoneten.
It is interesting to note that the REA itself can be expressed in a
similar form to \oneten, \mioneten\ and \defoneten\
\refmark{\rFMai,\rMAJ,\rSkld},
$$\CR K_1K_2 = K_2K_1 , \eqn\nREtwo
$$
in which $\CR$ is a matrix acting on $\Ctext\sp{N\sp2}\times\Ctext\sp{N\sp2}$
which can be expressed by a product of four ordinary $R$-matrices.
This form of RE will be important for the  commuting transfer
matrix for non-periodic boundary conditions of lattice models to be discussed
in section 6.
For concreteness let us write \nREtwo\ with full indices
$$ \CR_{ij\,uv;~i'j'\,u'v'}K_{i'j'}K_{u'v'}= K_{uv}K_{ij}.
$$
For the RE1 the explicit form of \CR\ is
$$ \CR_{ij\,uv;~IJ\,UV}= \Rt\sp{-1}_{iv\,i'v'}\Rt\sp{-1}_{jv'\,j'V}
R_{i'u\,Iu'}R_{j'u'\,JU}, \eqn\CROne
$$
and for the RE2
$$ \CR_{ij\,uv;~IJ\,UV}=(R\sp{t_2})\sp{-1}_{iv\,i'v'}
(\wt {R\sp{t_1\circ t_2}})\sp{-1}_{jv'\,j'V}
R_{i'u\,Iu'}\wt{R\sp{t_2}}_{j'u'\,JU}, \eqn\CRTwo
$$
One can also write down in a similar fashion
 the relations among $K\sp{(i)}$ and $K\sp{(j)}$ of
the extended REA.

Now one can require the consistency among various kinds of $q$-tensors.
For example an exchange relation between $K_{ij}$ and $V\sp{(1)}=\{v_k\}$
a $q$-vector reads
$$ \CR\sp{(1,2)}V_1\sp{(1)}K_2 = K_2V_1\sp{(1)}. \eqn\onetwoten
$$
The consistency with the quantum group coaction requires
$$ \CR\sp{(1,2)}T_1T_2(T\sp{-1})\sp t_{2'}= T_2(T\sp{-1})\sp t_{2'}T_1
\CR\sp{(1,2)}. \eqn\cononetwo
$$
An  explicit form of $\CR\sp{(1,2)}$ satisfying \cononetwo\ is given by
$$\CR\sp{(1,2)} = ((\wt R_{12'}\sp{-1})\sp{t_2})\sp{-1}(\wt R_{12})\sp{-1}.
\eqn\Rontw
$$
The exchange relation between $V\sp{(1)}$ and $K$ can be rewritten in
a simpler form
by using  the usual notation for the reflection equations
$$ (\wt R)\sp{-1}V\sp{(1)}_1K_2 = K_2(\wt R)\sp{-1}V\sp{(1)}_1. \eqn\KVeq
$$
Moreover there is a natural \lq\lq coaction'' of $K$ on $V\sp{(1)}$:
$$ \wt\delta:~~V\sp{(1)} \to KV\sp{(1)} =   {V'}\sp{(1)}. \eqn\Kcoact
$$
It is straightforward to verify that $ {V'}\sp{(1)}$ satisfies
\defoneten\ by using RE2 and \KVeq. Namely they are again
$q$-vectors. We might call this a \lq\lq contraction\rq\rq\ of a (1,1)
tensor with a (1,0) tensor (a $q$-vector).

One can require the consistency between the two kinds of rank two $q$-tensors,
$K$ for RE2 and $\bar K$ for RE1,
which reads
$$ \Rt\sp{-1}\bar K_1R\sp{t_1}K_2 = K_2\wt R\sp{-1}\bar K_1R\sp{t_1}.
\eqn\comontw
$$
Similarly to the above case \Kcoact, there is a natural \lq\lq coaction''
 $K$ on
$\bar V\sp{(2)} \sim \bar K$:
$$\wt\de:~~\bar V\sp{(2)} \to K\bar V\sp{(2)} =  K\bar K. \eqn\Kcoactone
$$
It is easy to show that $K\bar K$ satisfies RE1 by using \comontw.
As before, for the trivial element of the RE2 algebra, \ie $K=1$,
the above consistency
conditions among the $q$-tensors, \KVeq\ and \comontw, are trivially
satisfied.
The relation \Kcoactone\ may be considered again as the
\lq\lq contraction\rq\rq\
of $K$, a (1,1) tensor, and $\bar K$, a (2,0) tensor, to form
another (2,0) tensor.

One can decompose a $(1,1)$ tensor (\ie, the $K$-matrix of RE2) into a matrix
product
of $(1,0)$ tensor say $U\sp+$ and $(0,1)$ tensor say $U\sp-$,
$$ K = U\sp+U\sp-. \eqn\Kuurel
$$
The component $q$-tensors should  satisfy
$$RU\sp+_1U\sp+_2 = \al U\sp+_2U\sp+_1, \quad
 U\sp-_2U\sp-_1\Rt=\al U\sp-_1U\sp-_2, \eqn\Ruu
$$
in which $\al$ is one of the eigenvalues of $\hat R$ as before.
The quantum group coaction is given by
$$\delta:\quad U\sp+ \to TU\sp+,\quad U\sp- \to U\sp-T\sp{-1}. \eqn\Ucoact
$$
The consistency condition among them is
$$U\sp-_1\Rt U\sp+_2 = U\sp+_2U\sp-_1. \eqn\upluumin
$$
This decomposition was also introduced in \refmark\rFMai\ for the
ultra-localization
of the lattice Kac-Moody algebra.
 The above decomposition \Kuurel\ is definitely \lq invariant' under the
redefinition
$U\sp+ \to U\sp+S$ and $U\sp- \to S\sp{-1}U\sp-$.
  One can also rewrite \Kuurel\
as in \refmark\rFMai\ $K=U\sp+K_0U\sp-$, in which $K_0$ is an arbitrary
c-number matrix.
Then the above \lq gauge' transformation should be augmented by
$K_0\to S\sp{-1}K_0S$.

Generalization of these examples and discussion to higher rank tensors
is straightforward.

\chapter{Central Elements of REA:}

In this section we will discuss the central elements of (extended) REA
and/or of the general $q$-tensors. For concrete examples, see \qoncen\ and
\centwo\ in section 2. In the language of the $q$-tensors, the central
elements correspond to various invariants of the $q$-tensors.
In general, the central elements of an arbitrary YBA \FRTeq\ are not easily
known. The best known examples are the so-called quantum determinants for
certain YBA's.
Therefore the central elements for general $q$-tensors are not easily
calculable, except  for the ($n$,$n$) $q$-tensors.
In these cases, for which the (extended) REA is the best example, the
dependence on the YBA (or the quantum group) can be cancelled by choosing
certain combination of the covariant and contravariant indices.

Let us consider  an $N\times N$ matrix \CD\ which enjoys the following property
$$ {\rm Tr}(\CD M) = {\rm Tr}(\CD TMT\sp{-1}) , \eqn\cominvD
$$
in which $M$ is an arbitrary $N\times N$ matrix and $T$ is the matrix of
quantum group
generators associated with an $R$-matrix \FRTeq.
In other words, \CD defines an invariant trace with respect to the
(1,1) $q$-tensor (or the comodule  transformation of  the RE2).
In terms of \CD we can express the central elements of the REA
$$ c_n(K)= {\rm Tr}(\CD K\sp n), \eqn\cnK
$$
or more generally for the extended REA system
$$ c_n(K\sp{(j)})= {\rm Tr}(\CD (K\sp{(j)})\sp n), \quad 1\leq j \leq g, \quad
n: {\rm integer}. \eqn\cnj
$$
Namely they satisfy
$$ [\,c_n(K\sp{(i)}), K\sp{(i)}] = 0, \quad [\,c_n(K\sp{(i)}), K\sp{(j)}] = 0,
\qquad {\rm for}~ \forall i ,~ \forall  j. \eqn\centij
$$
Similarly we can define the central elements for $K\sp{(I)}$
as $c_n(K\sp{(I)})$ in which
$I$ is an ordered set of indices, for example $I=(i,j,k)$, $i>j>k>l$,
$$ [\,c_n(K\sp{(I)}), K\sp{(I)}] = 0, \quad [\,c_n(K\sp{(I)}), K\sp{(l)}] = 0.
 \eqn\centIl
$$

The explicit forms of \CD for some particular $R$-matrices have been known
for sometime \refmark\rFRT.
Here we give a simple derivation of the explicit form of \CD\ for an
arbitrarily given $R$-matrix, namely
$$ \CD = {\rm Tr}_2[\pe((R\sp{t_1})\sp{-1})\sp{t_1}]. \eqn\Dform
$$
In order that \cominvD\ should hold for any matrix $M$, \CD should satisfy
$$ \CD\sp t = T\sp t\CD\sp t(T\sp{-1})\sp t. \eqn\CDdef
$$
Starting from \FRTeq\ it is easy to get
$$ (R\sp{t_1})\sp{-1}T_2\sp{-1}T_1\sp t = T_1\sp tT_2\sp{-1}(R\sp{t_1})\sp{-1}.
\eqn\Rtinv
$$
By taking a sum over the second indices of the first and second spaces,
$T_2\sp{-1}$
and $T_1\sp t$ on the left hand side cancel to each other to give
$$ \CD\sp t = T\sp t\CD\sp t(T\sp{-1})\sp t,
$$
in which
$$\CD\sp t = {\rm Tr}_2[((R\sp{t_1})\sp{-1})\sp{t_2}\pe] .
$$
Namely we arrive at
$$\CD = {\rm Tr}_2[\pe((R\sp{t_1})\sp{-1})\sp{t_1}].
$$

Next we give a sketch of the proof that $c_n(K)$ in \cnK\ are central in the
RE2 algebra. By multiplying $K_2$ successively from the right to the RE2,
$$  R K_1 \Rt K_2 = K_2 R K_1 \Rt , \eqn\Retwo
$$
we get
$$  R K_1 \Rt K_2\sp n = K_2\sp n R K_1 \Rt,\qquad  n:~{\rm integer}
\eqn\retwon$$
which is rewritten as
$$ \Rt K_2RK_1\sp n = K_1\sp n\Rt K_2R,
$$
and further as
$$K_2RK_1\sp nR\sp{-1} = {\Rt}\sp{-1}K_1\sp n\Rt K_2. \eqn\krkn
$$
Since $T=R$ and $T=\Rt\sp{-1}$ satisfy the FRT relations \FRTeq\
we see that \CDdef\ implies,
in particular,
$$ {\rm Tr}_1(\CD_1RM_1R\sp{-1})= {\rm Tr}(\CD M)1_2, \quad
{\rm Tr}_1(\CD_1\Rt\sp{-1} M_1\Rt)= {\rm Tr}(\CD M)1_2,
$$
in which, as before, $M$ is an arbitrary matrix and $1_2$ is a unit matrix
in the second space.
Thus, by multiplying $\CD_1$ on both sides of \krkn\ and taking the
trace in the first space we obtain
$$K_2c_n(K) = c_n(K) K_2,
$$
or
$$[\,K, c_n(K)] = 0. \eqn\centcn
$$
It is straightforward to generalize the above centrality arguments
to the extended REA.
One only has to note that \Compp\ implies
$$R K_1\sp{(i)}R\sp{-1}(K\sp{(j)})\sp n_2 =
(K\sp{(j)})\sp n_2R\,K_1\sp{(i)}R\sp{-1}. \eqn\Comppn
$$
This corresponds to  \retwon\ above. It should be remarked that for higher
$n$, $c_n(K)$ is linearly dependent on lower $c_n(K)$'s
due to the (generalized)
Cayley-Hamilton theorem fot the matrix $K$ (see \refmark\rKuli).

Suppose we have a consistent set of a (1,1) and a (2,0) tensor (\ie, $K$
of RE2 and $\bar K$ of RE1) satisfying the condition \comontw.
Then  the central elements of the (1,1) tensor (RE2 algebra) commute
with  the (2,0) tensor ($\bar K$ of RE1)
$$ [\,c_n(K), \bar K] = 0. \eqn\cKcomK
$$
Here the following formula is useful
$$  \Rt\sp{-1}\bar K_1R\sp{t_1}K_2\sp n =
K_2\sp n\wt R\sp{-1}\bar K_1R\sp{t_1}.
$$

In a similar fashion we can show that the central elements of
a (1,1) tensor ($K$ of RE2) commute with the $q$-vector $V\sp{(1)}$ if
\KVeq\ is satisfied.

A certain class of YBA is known to have some additional quadratic relations
besides \FRTeq.
For example, for the YBA based on the orthogonal and symplectic
$R$-matrix there exists a c-number  matrix \CC
satisfying the relation \Ref\rTakha{\Takha}
$$ T\CC T\sp t = \CC\quad {\rm and}\quad T\sp t\CC T =\CC,
\quad \CC\sp2=\epsilon I, \eqn\CCprop
$$
in which $\epsilon=1$ for $B_n$ and $D_n$ whereas $\epsilon=-1$ for $C_n$.
In these cases the distinction between the covariant and the contravariant
indices becomes inessential and we can express the central element of
the RE1 algebra,
which is $n$-th order in $\bar K$, as
$$ \bar c_n(\bar K) = {\rm Tr}(\CC\sp t (K\sp{'})\sp{n-1}\bar K),
\quad K\sp{'}\equiv \bar K\CC.
\eqn\centone
$$
This indicates that \CC and \CD are related. In fact, we should have
$$\CD=\CC\sp{-1}\CC\sp t \quad {\rm or}\quad \CD =\CC\CC\sp t . \eqn\cdrel
$$
In the simplest case of the RE1 algebra for the $GL_q(2)$ $R$-matrix
discussed in section 2 \Ala, $\CC=\ep_q$  in \sol\ and
$\bar c_1$ and $\bar c_2$ above are essentially the same as $c_1$ and $c_2$
given in \qoncen.

\chapter{Commuting Transfer Matrices for Non-periodic Lattice}

In this section we discuss another important application of the REA,
the construction of commuting transfer matrices for non-periodic boundary
conditions on lattice statistical models, which was proposed in
\refmark{\rSklc} and further developed in
\refmark{\rMN,\rFMai}.
We have seen in section 2.1 that the usual \lq rapidity\rq-dependent
YBA (in this context the \lq spectral parameter\rq\ $\lambda$ might
be preferred to the \lq rapidity\rq\ $\theta$ but we stick to the convention)
with the comultiplication \coml\ gives a one parameter family of
commuting transfer matrices  associated with  periodic boundary conditions.
So, our starting point is the general \lq rapidity\rq-dependent RE
\REgen\ with the quantum group comodule property \twococat.
We need another ingredient, a \lq\lq dual\rq\rq\ reflection equation or a
\lq\lq dual\rq\rq\ $q$-tensor. The RE and its dual RE could be interpreted
as describing the \lq reflection\rq\ at the two ends of a line segment
corresponding to the finite lattice.

The essential ideas can be most easily explained in the case of $q$-vectors
 instead of the rank two $q$-tensors (the REA).
Let us start from the \lq rapidity\rq-dependent $q$-vector equation
\defoneten\ (which could be interpreted as the defining equations of a
Zamolodchikov algebra \refmark\rYZOW)
$$ R(\theta_1 -\theta_2)v(\theta_1)_1v(\theta_2)_2 =
\al v(\theta_2)_2v(\theta_1)_1.
\eqn\raptenone
$$
By applying the \lq quantum group coaction' $L$ times,
$L$ being the lattice size of
a lattice problem, we arrive at
$$\eqalign{& R(\theta_1-\theta_2)\left[(TT'\cdots T\sp{(L)})v(\theta_1)
\right]_1
\left[(TT'\cdots T\sp{(L)})v(\theta_2)\right]_2  \cr
&~~~~=\al
\left[(TT'\cdots T\sp{(L)})v(\theta_2)\right]_2\left[(TT'\cdots T\sp{(L)})
v(\theta_1)\right]_1
\cr} \eqn\RapL
$$
Next we introduce the \lq\lq dual\rq\rq\ $q$-vector by
$$ u\sp t(\theta_2)_2u\sp t(\theta_1)_1R(\theta_1-\theta_2)=
\al u\sp t(\theta_1)_1u\sp t(\theta_2)_2,
\eqn\dualvec
$$
which can be rewritten as
$$\al u\sp t(\theta_1)_1u\sp t(\theta_2)_2R\sp{-1}(\theta_1-\theta_2)
=  u\sp t(\theta_2)_2u\sp t(\theta_1)_1.
\eqn\Dvec
$$
The $R$-matrix dependence is cancelled by multiplying \RapL\ and \Dvec\
and we obtain the commuting transfer matrix
$t(\theta)$
$$[\,t(\theta_1), t(\theta_2)] = 0,\qquad t(\theta) =
u\sp t(\theta)(TT'\cdots T\sp{(L)})v(\theta),
\eqn\Rcomtr
$$
provided that $u(\theta)$ commutes with
$(TT'\cdots T\sp{(L)})v(\theta)$, component-wise.
Obviously this condition is satisfied  if the  dual equation
has a c-number solution.
However, to the best of our knowledge, the algebra of rank one $q$-tensors
 does not usually admit {\it non-trivial c-number solutions}
in contradistinction to the rank two $q$-tensors (the REA), which have a
variety of c-number solutions as exemplified in \refmark\rKSS.
This is why the REA is used for the derivation of commuting transfer matrices
for non-periodic lattices
\refmark{\rSklc,\rMN,\rFMai}.

The actual derivation of the commuting transfer matrix for the rank two
$q$-tensors (the REA) runs almost parallel with the rank one case if we
start from the rewritten form of the RE \nREtwo
$$\CR(\theta_1,\theta_2) K(\theta_1)_1K(\theta_2)_2 =
K(\theta_2)_2K(\theta_1)_1.  \eqn\RREtwo
$$
As is clear from the rank one case, the detailed form of $\CR$ is
inessential so long as \RREtwo\ has the covariance under the necessary
coaction
$$K(\theta)\to K_T(\theta)=
(TT'\cdots T\sp{(L)})K(\theta)(TT'\cdots T\sp{(L)})\sp\sigma,
\eqn\RKcom
$$
in which the anti-automorphism $\sigma$ is either the inverse or the transpose
according to the specific problem.
It should be noted that the \lq rapidity\rq\ of the $T$-matrix in the second
group (the $T\sp\sigma$ part) can be different from $\theta$.
In most studied cases, it is $-\theta$ corresponding to the change of the
\lq rapidity\rq\ after mirror reflection.
The dual rank two tensor is introduced  in a similar way to
the rank one case and is written symbolically as
$$J\sp t(\theta_2)_2J\sp t(\theta_1)_1\CR(\theta_1,\theta_2)=
J\sp t(\theta_1)_1J\sp t(\theta_2)_2.
\eqn\dualten
$$
We get a one parameter family of commuting transfer matrix
$t(\theta)$
$$[\,t(\theta_1), t(\theta_2)] = 0,\qquad t(\theta) =
{\rm Tr}\left[J\sp t(\theta)(TT'\cdots T\sp{(L)})K(\theta)
(TT'\cdots T\sp{(L)})\sp\sigma\right].
\eqn\RComtr
$$
Namely, the various c-number solutions of the (dual) RE specify the
possible non-periodic boundary conditions compatible with the integrability.
\chapter{Summary and Comments}
Many remarkable properties of the reflection equation algebras and their
extensions are elucidated from the view point of the covariance under
the  quantum group transformations or \lq\lq quantum group
tensors\rq\rq.
They include the generation of a series of new solutions by composing
known ones for the REA and for other $q$-tensors, the central elements of REA,
construction of integrable lattice models with  non-periodic boundary
conditions, etc.
We believe that the general framework introduced here will be quite useful
for actual problems related with the quantum groups or YBA.
To name some of them, the non-commutative differential geometry, deformation
of conformal field theory, quantum gravity and lattice version of Kac-Moody
algebra.

After finishing the manuscript we were informed that S.~Majid
\refmark\rMAJa\ had also come across with some of the examples of
$q$-tensors discussed in section 4. We thank him for the information.
We also thank E.~K.~Sklyanin and E.~Corrigan for useful comments.
P.~K. thanks to Yukawa Institute for Theoretical Physics, Kyoto University,
in which this work started. R.~S. thanks the Department of
Mathematical Sciences, Durham for hospitality.
\REF\rKulE{\KulE}

\refout
\end